\documentclass[10pt,onecolumn]{IEEEtran}

\usepackage{booktabs}
\usepackage{url}
\usepackage[justification=centering]{caption}
\usepackage{graphicx}
\usepackage{float}

\begin{document}

\title{Crisis Analytics: Big Data Driven Crisis Response}
\author{Junaid Qadir$^{1}$,
        Anwaar Ali$^{1}$, 
        Raihan ur Rasool$^{1,2}$,
        Andrej Zwitter$^{3}$,
        Arjuna Sathiaseelan$^{4}$,
        Jon Crowcroft$^{4}$\\
\normalsize $^{1}$School of EE and CS (SEECS), National University of Sciences and Technology (NUST), Pakistan\\
\normalsize $^{2}$King Faisal University (KFU), Kingdom of Saudi Arabia\\
\normalsize $^{3}$University of Groningen, Netherlands\\
\normalsize $^{3}$Computer Laboratory, University of Cambridge, United Kingdom\\
\vspace{1mm}\small \{junaid.qadir, 13mseeaali\}@seecs.edu.pk; rrasool@kfu.edu.sa; a.zwitter@rug.nl; 
\{arjuna.sathiaseelan, jon.crowcroft\}@cl.cam.ac.uk\\
}
\maketitle


\begin{abstract}

Disasters have long been a scourge for humanity. With the advances in technology (in terms of computing, communications, and the ability to process and analyze big data), our ability to respond to disasters is at an inflection point. There is great optimism that big data tools can be leveraged to process the large amounts of crisis-related data (in the form of user generated data in addition to the traditional humanitarian data) to provide an insight into the fast-changing situation and help drive an effective disaster response. This article introduces the history and the future of big crisis data analytics, along with a discussion on its promise, challenges, and pitfalls.

\end{abstract}

\section{Introduction}

In 2010, Haiti faced a catastrophic disaster: a devastating earthquake had struck the country's capital, Port-au-Prince. The human toll of this earthquake was immense---an estimated 100,000 to 300,000 people were dead, and the majority of Port-au-Prince had become homeless. Amongst the many in the need of help were two old women and a young girl of 7 who were stuck under the rubble of a collapsed roof of a supermarket in an area of Port-au-Prince known as Grand Goave. Fortunately, the humanitarian response came quick. Driven by information and communication technology, humanitarian aid was different this time. What followed heralded the cusp of a new era: the dawn of what is now known as \textit{digital humanitarianism} \cite{meier2014digital}.

Digital humanitarians used a number of emerging technologies---read-write web, mobile technology, social media, big data analytics, participatory mapping, and  crowdsourced translation---to catalyze an effective response.  A lot of crisis data---including short message service (SMS) from on-site victims, social media data from citizens, journalists, and aid organizations---was subsequently collected. Sifting through the voluminous ``\textit{big crisis data}'' to find information about the trapped girl and women was like the proverbial search for a needle in the haystack. Digital humanitarians successfully tackled this challenge using various techniques (most notably by crowdsourcing data and mapping it). The platform through which this crowdsourcing was managed for the Haiti earthquake was Ushahidi. Ushahidi---which is a free and open source software---provides a mobile-based platform for developing ``\textit{crowd maps}'' through the process of collecting, visualizing, and mapping citizen-supplied data. In January 2010, Ushahidi was used by Haitians to provide 3,584 reports, which were viewed approximately 500,000 times---this crowdsourced information, which was then visualized on a map, allowed Haitians to report their location and needs. Based on the gathered intelligence, the digital humanitarians alerted an onside aid unit (the 22 Marine Expeditionary Unit) towards the need of help at  Grand Goave. The 22MEU team was able to evacuate the 7-year old girl and the two old women from the rubble.

It has been known for quite some time that a critical component of effective emergency response is information dissemination and accessibility during disasters. In the hyperconnected modern world, crisis-affected communities will often resort first to technology following a sudden disaster by calling, messaging, emailing, or tweeting to connect with their family and report the incident to the government and aid organizations. This data---comprising updates about the changing situation; the need of immediate medical attention; the locations of makeshift aid stations---presents an invaluable insight into the dynamic situation on the crisis site. This data is useful in two broad contexts: (1) it helps the international aid community in ascertaining the crisis situation; (2) crucially, it acts as an information source for the crisis affected. 

The 2010 Haiti earthquake response will be remembered as a landmark moment in Internet's evolution, when digital technology (in the form of mobile and online communication) became an integral part of disaster response in an emergent way. In the Haiti crisis, thousands of citizens around the world collaborated in volunteer and technical communities (V\&TCs) to help make sense of a large-scale calamity. This led to the emergence of a new kind of distributed intelligence: the new era of \textit{Disaster Response 2.0} \cite{crowley2011disaster} had arrived.

\section{What is Big Data? Why use it for Crisis Analytics?}

To be sure, the use of mapping technology to combat crisis is not new. A classic example of Crisis Mapping Analytics is John Snow's Cholera Map.  Snow studied the severe outbreak of cholera in 1854 near the Broad Street in London, England. Contrary to the prevailing mindset (that believed that cholera was spread through polluted air or ``miasma''), Snow showed---through the ingenious use of spatial analytics (comprising mapping and detailed statistical analysis)---that the cholera cases were all clustered around the pump in a particular street, and that contaminated water, not air, spread cholera. We can also turn to  Florence Nightingale---a nurse assigned to the old Barrack Hospital in Scutari during the Crimean War in 1850s---for another vintage example of data-based crisis analytics. Nightingale combined data analytics and striking visualizations to highlight the importance of proper healthcare and hygiene for checking the spread of disease amongst soldiers. In the work of Snow and Nightingale, we already see the vestiges of crisis analytics: they were already testing for spatial effects (such as autocorrelation; clustering/ dispersion) and testing hypotheses (about proposed correlations and relationships). 

\vspace{1mm}
\textit{It is worth asking then: what exactly is new about big crisis data analytics?} 

\vspace{1mm}
We discuss the technologies enabling the new field of big crisis data analytics later, but before we get to that, let us first highlight what big data is. 

\subsection{What is Big Data?}

In the modern world we are inundated with massive amounts of data \cite{james2011big}. The data deluge is increasing exponentially due to the increased digitization of modern life and the commoditization of data collection with the advancement in digital technology. Companies such as Facebook, Google, Twitter, Yahoo, and Microsoft routinely deal with petabytes of data on a daily basis. \textit{``Big data''} refers to our emerging ability to collect, process, and analyze massive sets of largely unstructured data from multiple sources to find previously inaccessible insights. Despite the name \textit{big} data, large volume is only one of the big data essentials. To be sure, there is no fixed boundary differentiating big data from small data: big data typically refers to large datasets that cannot be processed on a single state-of-the-art machine using traditional database and data warehousing/ analytics tools (since the data is too large, too fast, or unstructured). 





\subsection{Sources of Big Crisis Data}

\begin{figure}[t]
\begin{center}
\includegraphics[width=6in]{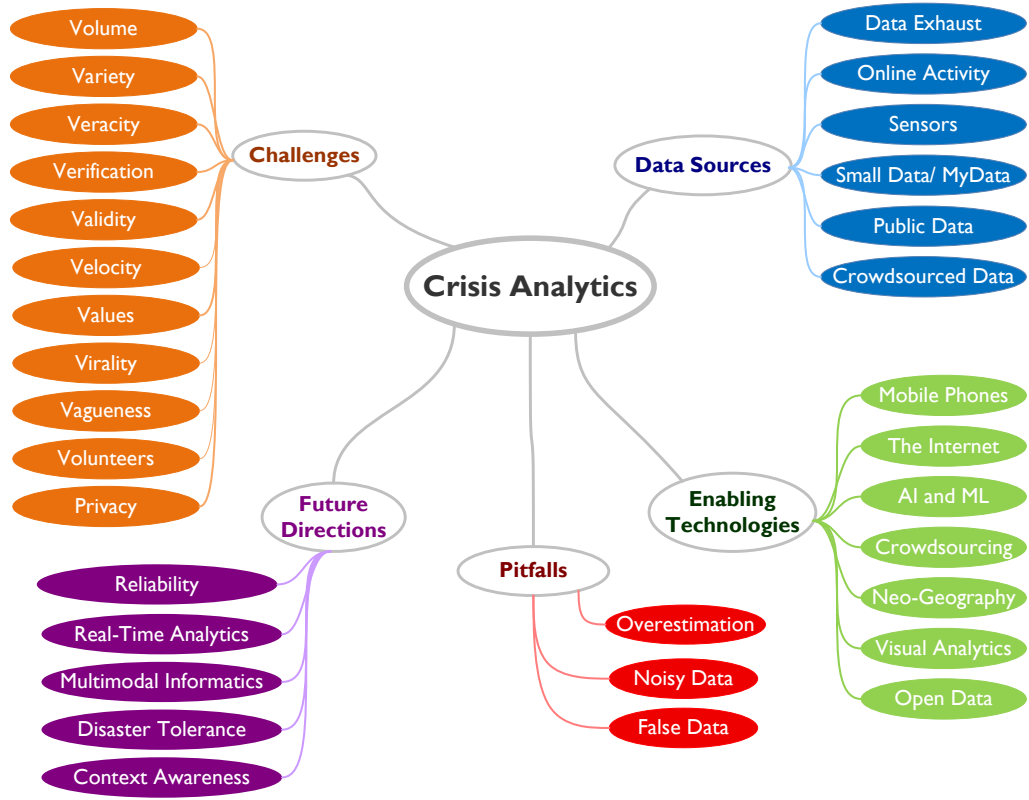}
\caption{Taxonomy of Crisis Analytics} 
\label{fig:taxonomy}
\end{center}
\end{figure}

As depicted in a detailed crisis analytics taxonomy shown in Figure \ref{fig:taxonomy}, there are six important sources of big crisis data. 


\begin{enumerate}

\vspace{1mm}
\item \textit{Data exhaust} refers to the digital trail that we etch behind as we go about performing our everyday online activities with digital devices. The most important example of data exhaust for big crisis data analytics is the mobile ``\textit{call detail records}'' (CDRs), which are generated by mobile telecom companies to capture various details related to any call made over their network.  Data exhaust also includes transaction data (e.g., banking records and credit card history) and usage data (e.g., access logs). Most of the data exhaust is owned by private organizations (such as mobile service operators) where it used mostly in-house for troubleshooting; data exhaust is seldom shared publicly due to legal and privacy concerns.


\vspace{1mm}
\item \textit{Online activity} encompasses all types of user generated data on the Internet (e.g., emails, SMS, blogs, comments); search activity using a search engine (such as Google search queries); and activities on social networks (such as Facebook comments, Google+ posts, and Twitter tweets). It has been shown in literature that online activities on different platforms can provide unique insights to crisis development: as an example, the short message services Twitter and SMS are used differently in crisis situations---SMS is used mostly on the ground by the affected community, while Twitter is used mostly by the international aid community \cite{munro2012short}. The advantage of online data is that it is often publicly available, and thus it is heavily used by academics in big crisis data research.


\vspace{1mm}
\item \textit{Sensing technologies} use various cyber-physical sensing systems---such as ground, aerial, and marine vehicles; mobile phones; wireless sensor nodes---to \textit{actively} gather information about environmental conditions. There are a number of sensing technologies such as (1) \textit{remote sensing} (in which a satellite or high-flying aircraft scans the earth in order to obtain information about it); (2) \textit{networked sensing} (in which sensors can perform sensing and can communicate with each other---as in wireless sensor networks); and (3) \textit{participatory sensing} (in which everyday entities---such as mobile phones, buses, etc.---are fit with sensors). With the emergence of the Internet of Things (IoT) architecture, it is anticipated that sensor data will become one of the biggest sources of big crisis data. Sensing data is usually (but not always) publicly available. 


\vspace{1mm}
\item  \textit{Small data and MyData}:  With big data, the scope of sampling and analysis can be vastly dissimilar (e.g., the unit of sampling is at the individual level, while the unit of analysis is at the country level), but with ``\textit{small data}'', the unit of analysis is similarly scoped as the unit of sampling. When the unit of sampling and analysis is a single person, we call such personal-data-based analysis ``\emph{MyData}''. There is emerging interest in using small data and MyData for personalized solutions, focused on applications like health (e.g., Cornell's \textit{mhealth} project led by Deborah Estrin) and sustainable development (e.g., the Small Data lab at the United Nations University). Today individuals rarely own, or even have access to, all of their personal data; but this has started to change (e.g., some hospitals now make individual medical records data accessible to patients). 




\vspace{1mm}
\item  A lot of \textit{public-related data}---that can be very valuable in the case of a crisis---is already being collected by various public/ governmental/ or municipal offices. This includes census data, birth and death certificates, and other types of personal and socio-economic data. Typically, such data has been painstakingly collected using paper-based traditional survey methods. In recent times, advances in digital technology have led people to develop mobile-phone-based data-collection tools that can easily collect, aggregate, and analyze data. Various open-source tools such as the Open Data Kit (ODK) make it trivial for such data to be collected. While public-related data is not always publicly accessible, increasingly governments are adopting the Open Data trend to open up public-related data.


\vspace{1mm}
\item Finally, the method of \textit{crowdsourcing} is an active data collection method in which applications actively involve a wide user base to solicit their knowledge about particular topics or events. Crowdsourcing combines a) digital technology, b) human skills, and c) human generosity and utilizes the cognitive surplus of \textit{digital human samaritans}---the volunteer open-source coders; the citizens who provide data, or help complete a task---to create a volunteer workforce that can be put to work on large global projects. Crowdsourced data is usually publicly available and is widely used by big crisis data practitioners.  

\end{enumerate}

\subsection{Big Crisis Data Analytics}

 

Big data technologies can be divided into two broad fields: (1) \textit{big data engineering}---which deals with the problems of storing and processing big data using new kinds of database technology (such as NoSQL) and scale-out computing (such as Hadoop and Spark); and (2) \textit{big data analytics}---which uses ``\textit{data science}'' (a combination of skills from the fields of computer, statistics, signal processing, data mining, machine learning, and visualization) to mine insights from big data. In this article, our focus is on the category of big data analytics---particularly, in the application of big data analytics on crisis data for facilitating humanitarian action. 



The aim of \textit{big crisis data analytics} is to leverage big data techniques---comprising techniques from artificial intelligence (AI), machine learning (ML), and data analytics---along with digital platforms (such as mobile phones/ Internet) for efficient humanitarian response to different crises. There are many thematic \textit{applications of big crisis data analytics} including (1) data-driven digital epidemiology (in which public health research is conducted using CDRs and social media) \cite{salathe2012digital}; (2) population surveillance and urban analytics \cite{boulos2011crowdsourcing} (in which big crisis data is used for tracking the movement of crisis-affected population as the crisis unfolds); (3) crisis informatics and sociology \cite{palen2007crisis} (in which data, along with participatory mapping and crowdsourcing technology, is used for analyzing the sociological behavior of the affected community through behavioral inference and ``\textit{reality mining}'').

\section{Big Crisis Data Analytics: Enabling Technologies}
 
\subsection{Mobile Phones} 


The rapid adoption of mobile technology has been unprecedented. Smartphones are rapidly becoming the central computer and communication devices in the lives of people around the world. Modern phones are not restricted to only making and receiving calls---current off-the-shelf smartphones can be used to detect, among other things, physical activity (via accelerometers); speech and auditory context (via microphones); location (via GPS) and co-location with others (via Bluetooth and GPS). This transforms the modern crisis response since modern smartphones can now act as general-purpose sensors and individuals can directly engage in the disaster response activities through cloud-, crowd-, and SMS-based technologies. This participatory trend in which the aid efforts are centered on and driven by people---and the fact that aid workers have to work with large amounts of diverse data---makes modern disaster response totally different from traditional approaches.

Mobile phone technology is ubiquitously deployed, both in developed countries as well as in underdeveloped countries. CDR-based mobile analytics presents a great opportunity to obtain insights (at a very lost cost) about mobility patterns, traffic information, and sociological networks---information that can be profitably utilized during various stages of disaster response (e.g., in epidemic control, and in tracking population dynamics).  CDRs have been used by digital humanitarians  during various crises (such as the non-profit \textit{FlowMinder}'s work with anonymous mobile operator data during the Haiti earthquake to follow the massive population displacements) to not only point out the current locations of populations, but also predict their future trajectory \cite{salathe2012digital}.

\subsection{The Internet, Open Source, and Open Data}

In a wide variety of fields, the new Internet-based economy is spawning a paradigm shift in how institutions work. The ``\textit{open source}'' culture (the paradigm underlying Internet projects such as Linux and Wikipedia) has ushered in a new era that relies more on collaboration and volunteerism (than on formal organizations). Instead of a rigid structure constrained by scarcity of human resources, the new paradigm is driven by abundance and cognitive surplus (due to technology-driven pooling of volunteer human resources). This open source trend is now also visible in humanitarian development in various manifestations such as digital humanitarianism, user generated information, participatory mapping, volunteered geographic information, open-source software (such as \textit{OpenStreetMap}) and open data. Many V\&TCs have exploited these open standards to link data from disparate sources and create \textit{mashups} (which are defined as a web page/ application that uses or combines data or functionality from multiple existing sources to create new services).

Another important trend emerging in the modern era is the ``\textit{Open Data}'' that has resulted in unprecedented commoditization and opening up of data. A number of countries from allover the world (more than 40 countries in 2015) have established open data initiatives to open numerous kinds of datasets to public for greater transparency. Open data can also lead to improved governance through the involvement of public and better collaboration between public and private organizations. As an example, in the aftermath of the Haiti crisis, volunteers across the world cobbled together data from various sources---including data from satellite maps and mobile companies along with information about health facilities from the maps of the World Health Organization, and police facilities from the Pacific Disaster Center---and plotted them on open-source platforms such as the OpenStreetMap. 
The importance of OpenStreetMap can be gauged from the fact that soon after the earthquake, OpenStreetMap had become the de facto source of Haiti map data for most of the United Nations (UN) agencies.

\subsection{Neo-Geography and Visual Analytics}

Visual analytics is an exploratory data analysis tool that aims to support analytical reasoning through visual interfaces. With geographic data being critical for crisis response, the ability to visualize and map geographic data becomes indispensable. Sophisticated geographic maps are not new: indeed, geography and cartography have a long history, with high-quality maps available for hundreds of years. But with changing times, these fields have advanced. Neo-geography (``new geography'') prominently uses the Web 2.0's democratic approach of participatory mapping along with crowdsourced user-generated volunteered geographic content \cite{turner2006introduction}. Unlike the maps of yesteryears, with neo-cartography (``new cartography''), the data is not controlled by an elite few.  The highly-popular free and open-source software OpenStreetMap presents a concrete participating mapping example embodying these trends in which geographical data is collected from volunteers. 

Big crisis data visual analytics aims for creating images, maps, interactive animations that enhance big data analytics by (1) exploiting the human ability to recognize erstwhile hidden patterns when data is visualized; (2) automated processing of images and videos (as in crisis maps) by combining computer vision and geospatial techniques. A powerful example of these trends is the recent emergence of the cross-disciplinary field of \textit{``crisis mapping''} \cite{ziemke2012crisis}. With the ability to acquire up-to-date crisis-site imagery through crowdsourcing and the use of cyber-physical systems, digital humanitarians can have a virtual bird's-eye view to guide the response efforts (the sense of empowerment being akin to having one's own fleet of private helicopters, each with a dedicated pilot \cite{meier2014digital}). The up-to-date maps provide effective situational awareness in time and space by providing a dynamic view to guide the response efforts.

\subsection{Leveraging the Wisdom and the Generosity of the Crowd}

Broadly speaking, there are only a few ways we can go about problem solving or predicting something: (1) experts, (2) crowds, and (3) machines (working on algorithms; or learning from data). While experts possess valuable experiences and insights, they may also suffer from biases. The benefit of crowds accrues from its diversity: it is typically the case that due to a phenomenon known as ``the wisdom of the crowds'' \cite{surowiecki2005wisdom}, the collective opinion of a group of diverse individuals is better than, or at least as good as, the opinion of experts. Crowds can be useful in disaster response in at least two different ways: firstly, crowdsourcing, in which disaster data is gathered from a broad set of users and locations \cite{boulos2011crowdsourcing}; and secondly, crowdcomputing, in which  crowds help process and analyze the data through collaboratively solving ``microtasks'' \cite{meier2014digital}.

\vspace{1mm}
\subsubsection{Crowdsourcing}

Crowdsourcing is the outsourcing of a job traditionally performed by a designated agent (usually an employee) to an undefined---generally a large group of people---in the form of an open call. In essence, crowdsourcing is the application of the open-source principles (used to develop products such as Linux, Wikipedia, etc.) to the fields outside of software. Crowdsourcing has been used in the context of disaster response in multiple ways \cite{munro2013crowdsourcing}: including crowdsearching, microtasking, citizen science, rapid translation, data cleaning and verification, developing ML classifiers, and election monitoring \cite{boulos2011crowdsourcing}.

\vspace{1mm}
\subsubsection{Crowdcomputing}

Crowdcomputing is a technique that utilizes crowds for solving complex problems. A notable early use of crowdcomputing was the use of crowdsearching by MIT's team at the 2009 DARPA Network Challenge. The MIT's team solved a time-critical problem in the least time using an integration of social networking, the Internet, and some clever incentives to foment crowd collaboration. In contemporary times, a number of ``\textit{microtasking}'' platforms have emerged as smarter ways of crowdsearching. A familiar example is ``Amazon Mechanical Turk'', the commercial microtasking platform that allows users to submit tasks of a large job (that is too large for a single person or small team to perform) for distribution to a global crowd of volunteers (who are remunerated in return for performing these microtasks). A number of free and open-source microtasking platforms have also been developed, including generic microtasking platforms such as \textit{CrowdCrafting}---which was used by Digital Humanitarian Network (DHN) volunteers in response to Typhoon Pablo in the Philippines---as well as humanitarian-response-focused platforms such as \textit{MicroMappers}. MicroMappers, developed at the Qatar Computing Research Institute (QCRI), was conceived as a fully customized microtasking platform for humanitarian response---a platform that would be on standby and available within minutes of the DHN being activated. MicroMappers can facilitate the microtasking of translatio, and classification of online user-generated multimedia content (in various formats such as text, images, videos) through tagging. 



\subsection{Artificial Intelligence and Machine Learning}

AI and ML can be used to facilitate crisis response through automated problem solving. In many cases, algorithms (traditional or AI-based) can be formulated to attack a given problem. However, for many non-trivial problems, such exact algorithms are not known and a ML-based approach---based on learning from data---is more appropriate. 
 
\vspace{1mm}
\subsubsection{Tasks performed by AI/ ML algorithms}

The various \textit{big crisis analytics tasks} that can be performed on big crisis data are (a) discovery tasks---e.g., clustering (to obtain natural groupings), outlier detection (to detect any anomalies), or affinity/ correlation analysis (to detect co-occurrence patterns); and (b) various predictive tasks---such as classification (in which a category is predicted), regression (in which value of a continuous variable is predicted), and finally recommendation (in which some preference is predicted).

\vspace{1mm}
\subsubsection{Types of ML algorithms}

Broadly speaking, we can categorize ML algorithms into two types: \textit{supervised} and \textit{unsupervised}.

\vspace{1mm}
\textit{Supervised learning} uses techniques to infer a classifier function that generalizes a relationship between the input and the output from the given training data (that contains labeled examples of the data and the corresponding correct category). A number of supervised learning techniques exist including the popular and easy to use techniques of support vector machines (SVM) and random forests. Supervised learning works very well when labeled data is available but can suffer from \textit{``overfitting''} (when the learnt classifier is tightly coupled with the training data and cannot be reused when the setting changes). 


\textit{Unsupervised learning} techniques do not depend on the availability of training data, and can automatically learn features and find hidden structure in unlabeled data. Common examples of unsupervised learning techniques are the K-means and expectation-maximization (EM) clustering algorithms.  Unsupervised learning is becoming increasingly important in the big data era to alleviate the inconvenience of handcrafting features and providing labeled examples. In this regard, the emerging field of \textit{deep learning} holds a lot of promise.

\vspace{1mm}
\subsubsection{AI/ ML Domain Areas Relevant for Crisis Informatics}

While there can be a number of AI/ ML applications for crisis informatics, we discuss two important domain areas. 

\vspace{1mm}
\paragraph{Computational linguistics}

The most important application of AI-based computational linguistics for big crisis data analytics is automated analysis of social media using {sentiment analysis} and \emph{opinion mining} \cite{imran2015processing}. While there have been many advances in the field of natural language processing (NLP), applying NLP for crisis informatics is still non-trivial. 
Computational-linguistics-based AI techniques have been utilized extensively in many previous humanitarian responses. As an example, the \textit{CrowdFlower} platform, which provides crowdsourced translation services, was used in the response to the Haiti crisis (by the Mission 4636 project) and to the Pakistani flood crisis (by the PakReports project). 



\vspace{1mm}
\paragraph{Computer Vision}


There is a lot of interest in using AI-based computer vision techniques for automated analysis using aerial images (obtained via unmanned aerial vehicles (UAVs) or satellites). This interest has received a further fillip through various initiatives (such as the  ``Imagery to the Crowd'' initiative of the US government) that are aiming to democratize access to high-resolution commercial satellite images. There are various projects focused on automated analysis of global high-resolution satellite imagery, such as the computer vision projects at the European Commission's Joint Research Center (JRC)). Researchers can also use \textit{Tomnod}, the DigitalGlobe-owned microtasking platform exclusively focused on microtasking satellite imagery, for computer vision applications. 






\vspace{1mm}
\subsubsection{Interfacing Human and Artificial Intelligence}


The most potent approach to big crisis data analytics will leverage both the human and machine intelligence. This hybrid approach can exploit the complementary competencies of humans and machines---humans are better at finding patterns and codifying unstructured information while machines can process data faster---to simultaneously improve the pace and accuracy of crisis analytics. A promising direction in this regard is to use ML and to crowdsource the classifier generation to crowds (using microtasking-based crowdcomputing approaches). This is the underlying principle behind the AI for disaster response (AIDR) platform \cite{imran2014aidr} that is designed to make it trivial to generate new ML classifiers. AIDR can be used to automatically classify tweets (after it has been taught initially through the generation of the classifier). Since with automated ML, the possibility of erroneous classification is always present, AIDR also provides a confidence level with its auto-tags (which nicely allows integration of human and machine intelligence; e.g., by bringing the human into the loop when the machine's confidence level is low).

\section{Why Big Crisis Data Analytics is Challenging?}


\subsection{The ``Vexing Vs''}

The technical challenges associated with processing big data have traditionally been summarized using the four Vs: (1) \emph{volume} (large amounts of data that cannot be processed on a single machine or with traditional database tools); (2) \emph{variety} (data in many formats: structured, semi-structured, and unstructured---with mostly the data being unstructured); (3) \emph{velocity} (streaming data, with milliseconds to seconds to respond); and (4) \emph{veracity} (uncertainty in the data being ``true'' or consistent with reality). With big crisis data (such as social media), the list of ``vexing Vs'' is even longer \cite{castillo2016bigcrisisdata}: we also have to deal with (5) \emph{vagueness} (dealing with natural language); (6) \emph{virality} (ensuring that false information and rumors are not propagated); (7) \emph{volunteers} (motivating and coordinating digital volunteers); (8) \emph{validity} (mitigating the biases and pitfalls of social media); (9) \emph{values} (ensuring privacy and ethical use of crisis data); and finally, (10) \emph{visualization} (how to best visualize big crisis data such as crisis maps). For the sake of brevity, we will focus only on a subset of these challenges; more details can be seen in \cite{castillo2016bigcrisisdata}.

\subsection{Policy Challenges}

\vspace{1mm}
\subsubsection{Ensuring Privacy and Preventing Abuse}

The field of big data promises great opportunities but also entails some great risks of abuse and misuse. With big crisis data, there is always the danger of the wrong people getting hold of sensitive data---something that can easily lead to disastrous consequences. The development of appropriate policy can help manage this dilemma between the opportunities and risks of big data. Some of the big questions that big data policies should address are: (1) what data to open up? (2) who should be able to access which data? (3) which data should be publicly accessible? (4) how can the data be used, reused, repurposed, and linked? The devised policies must also include prescriptive steps that ensure that data is used ethically (and not misused by malevolent actors and crisis profiteers). In particular, we should take steps to ensure that crisis victims do not expose themselves or others to further harm unwittingly (e.g., in countries beset with a civil war, or sectarian violence, a request for help with personal information may also be used maliciously by malevolent actors for violent purposes).

\vspace{1mm}
\subsubsection{Ethical Big Crisis Data Analytics}

It is also important that the big crisis data and the digital humanitarian communities emphasize value-based and ethical humanitarian service. In this regard, these communities can leverage the collective knowledge of the existing humanitarian organizations available in the form of the \textit{``humanitarian principles''}\footnote{\url{https://docs.unocha.org/sites/dms/Documents/OOM_HumPrinciple_English.pdf}} that define a set of universal principles for humanitarian action based on international humanitarian law. These principles are widely accepted by humanitarian actors and are even binding for the UN agencies. The guiding humanitarian principles are: (1) \textit{humanity}: the humanitarian imperative comes first---aid has to be given in accordance to need. The purpose of humanitarian action is to protect life and health and ensure respect for human beings; (2) \textit{neutrality}: the humanitarian actors must not take sides in hostilities or engage in controversies of a political, racial, religious or ideological nature; (3) \textit{impartiality}: aid should be delivered without discrimination as to nationality, race, religious beliefs, class or political opinions; and (4)  \textit{independence}: the humanitarian action must be autonomous from the political, economic, military or other objectives that any actor may hold with regard to areas where humanitarian action is being implemented. The big crisis data analytics community also needs to adopt these, or similar, principles to guide their own work.

\section{Pitfalls with Big Crisis Data Analytics}

\subsection{Caveat Emptor: Beware of the Big Noise}

If big sized data was not challenging enough, crisis analysts have to deal with another formidable challenge: big false data. The presence of false data dilutes the signal to noise ratio, making the task of finding the right information at the right time even more challenging. This problem of big noise is particularly problematic for crowdsourced data in which the noise may be injected intentionally or unintentionally. Intentional sources of noise may come from pranksters or more sinisterly through cyber-attacks (this is particularly a risk during man-inflicted disasters, such as coordinated terrorist attacks). Unintentional sources of noise also creep into disaster data (e.g., through the spreading of false rumors on social networks; or through the circulation of stale information about some time-critical matter). The data may also be false due to bias. Much like the ``dog that didn't bark'' that tipped off Sherlock Holmes in one of his investigations, the data that is not captured is sometimes more important than what was captured. This sampling bias is always present in social media and must be investigated using sound statistical analysis (the need of which is not obviated due to the large size of data). As an example of the inherent bias in big data, we note that the Google Flu Tracker overestimated the size of the 2013 influenza pandemic by 50\%, and predicted double the amount of flu-related doctor visits. 




\subsection{Big Data/ Social Media/ Technology is not a Panacea}

While big data techniques (including those that leverage social media analytics) can accelerate and enhance the efficiency of the emergency response efforts, big crisis data analytics is but one cog in the entire crisis informatics ecosystem, and it cannot work in isolation. For instance, it has been reported that while Haiti response did benefit from big data efforts (crowdsourcing/ crisis mapping), more importantly, the response benefited most from an old technology (radio) and relied on the Haiti diaspora. With many stakeholders in the crisis response ecosystem, there is a need of harmonic smooth handover of responsibilities and for the linkage between the relief, rehabilitation and development (LRRD) processes. It is also worth emphasizing that technology is not a panacea and can only amplify the output of a project based on human intent and goals---it is thus the humans who are at the center of the digital humanitarianism universe \cite{meier2014digital}. 


\section{Future Directions For Big Crisis Data Analytics}

\subsection{Real-Time Crisis Analytics}

During disaster response, timely action is a matter of life and death. To complicate matters, accessing stale data may not be just harmless, it may even be harmful. This motivates the problem: \textit{how to effectively filter information from an overwhelming avalanche of big crisis data in real time?} The development of real-time big data analytics capability has many applications that can enhance disaster response. It is often the case that crises have cascading stages in which one problem may trigger other problems (e.g., a tsunami approaching a nuclear power plant). In such scenarios, real-time analytics can help prioritize the most urgent problems in a timely fashion before any further damage is caused. Critical information can also be broadcasted to the public (e.g., using radio technology or through SMS) so that any preventable follow-up hazards are thwarted. In addition, real-time analytics can assist aid organizations and first responders to coordinate with other stakeholders.



\subsection{Secure, Reliable, Disaster-Tolerant Crisis Analytics}

Given the gravity of crisis situations, it is important to ensure that big crisis data systems are highly disaster tolerant (in terms of reliability, availability, and security). This requires the incorporation of foolproof mechanisms to withstand extremely adverse conditions (such as flooding, earthquakes, physical damage to infrastructure, and power outages) and the usage of redundancy as well as diversity at all the levels (such as in the computing, storage, communications, sensing, and power components). Special attention should be paid to the power issue: many on-site communication/ computation devices (such as smartphones) continuously need battery charging or will otherwise cease to function. A combination of mobile technology and cloud computing (in the form of \textit{mobile cloud computing}) can be useful here as cloud computing naturally complements big data technologies and is well suited for reliable storage and analysis of big data. Since a system is only as secure as its weakest link, the focus on security should be all encompassing (and cover both the core cloud architecture as well as the mobile edge devices). It is also important to ensure proper authentication, authorization, and accounting: especially when providing remote access to resources (in the case of tele-operations such as tele-health) and sensitive data (such as the location and identification of affected people). 



\subsection{AI-Based Predictive/ Context-Aware Crisis Analytics}


Most of the work on crisis analytics has been hindsight focused, including works on \textit{Descriptive Analytics} (to answer ``what happened or is happening?'') or \textit{Diagnostic Analytics} (to answer ``why did it happen?''). Relatively few works (such as \cite{salathe2012digital},  \cite{perry2013predictive}) have focused on forward-looking analytics such as \textit{Predictive Analytics} (to answer ``what will happen?'').  With predictive crisis analytics, we can obtain advance information on developing disasters in an easy-to-understand and reliable way, such that the affected community can accept and act on the data. With the advances in AI, and the widespread availability of Internet and GPS with smartphones, it is possible to utilize predictive analytics to push out \textit{context-aware personalized crisis information} (e.g., using AI for automated processing and SMS for dissemination) to the crisis prone and the crisis affected. Consider the example of an earthquake: a simple application could be to send out notifications according to the information about the earthquake's epicenter and the location of the user; in more advanced applications, other sources of data (such as information from a user's social network) can be integrated to create applications that leverage knowledge about the user's location and friends to appropriately push messages. Facebook already offers a similar social networking based application called ``\textit{Safety Check}'' that users can use to inform and know about the safety of their friends and family after a disaster. For phones with active users, appropriate escape routes could also be pushed out to users using facilities such as the \textit{Google Crisis Response}.


\subsection{Multimodal Big Crisis Data Informatics}


Big crisis data incorporates a large number of data sources (that may be providing diverse kinds of data). Each of these big crisis data sources provides a unique (but necessarily incomplete) window to on-site developments. As popularized by John Godfrey Saxe's poem on ``the blind men and the elephant'', different experiences of the same reality can lead to different perceptions. This poem tells a story of the  six blind men who went to ``see'' an elephant and ended up in complete disagreement: the six blind men ``disputed loud and long, each in his own opinion; exceeding stiff and strong, though each was partly in the right, and all were in the wrong!''. This highlights the dangers of siloed big crisis data analytics on the fragmented data. To obtain more complete understanding of the reality of the crisis, it becomes important that we tackle the challenging task of reconciling and combining the various distinct modalities of crisis information (such as text, images, speech, video, maps, crowdsourced data, and formal reports). This field of multimodal big crisis data analytics defines the real frontier of research in big crisis data analytics and promises to be a fertile area of future research in the overall field of crisis informatics.

\bibliographystyle{IEEEtran}
\bibliography{bib-file}

\end{document}